\documentstyle[psfig,epsfig,12pt]{article}
\begin{document}
\pagestyle{plain}

\centerline{\bf SOCIAL PERCOLATION ON} 
\centerline{\bf INHOMOGENEOUS SPANNING NETWORK}

\begin{center}
Abhijit Kar Gupta$^\dag$\footnote{Postdoctoral 
fellow at: {\it Raman Research Institute, Bangalore-560 080, India}}
 and Dietrich Stauffer$^\ddag$
\end{center}

\begin{center}
{\it Institute for Theoretical Physics, Cologne University,\\
50923 K\"oln, Germany}\\
{\it E-mail:}~$^\dag${\it abhi@rri.ernet.in}, 
$^\ddag${\it stauffer@thp.uni-koeln.de}
\end{center}

\vskip 0.5 in

\abstract{The {\it Social Percolation} model recently 
proposed by Solomon {\em et al.} \cite{solo}
is studied on the Ising correlated inhomogeneous network. The
dynamics in this is studied so as to understand the role of correlations
in the social structure. 
Thus the possible role of the structural social connectivity is examined.}

\vskip 0.3 in


\vskip 0.5 in

In recent times, there has been a great interest in applications of Statistical
Physics in Social Science \cite{rev,soc}. These works are wide ranged; from the 
stock market data analysis to the `microscopic' models to understand 
the social dynamics. 
The study in social dynamics is concerned with the concept of 
information propagation and thus the formation of opinion in a complex social 
network. 
Various approaches have been developed in this respect.
A class of social impact models based on {\it social impact} theory has 
been dealt with in ref.\cite{impact}. 
One of the main premises in such models are
the interactions among the social agents and groups; the agents are 
influenced by one another while taking decision. 
Another approach is the 
assumption that the agents take decisions of their own (not influenced by each other), 
only the spreading of information is necessary through some kind of contact process, the 
philosophy of epidemiology (see a brief review \cite{briefreview}). 
The primary interest is to study through simple models how a complex process of
opinion formation (and regulation) takes place among the agents in a social space when the 
information is shared locally by `word of mouth' (`nearest neighbour' 
connection may be). This is
prior to a domination by powerful media ({\em e.g.}, TV, Newspapers, 
Internet) which may influence the agents (people) quite homogeneously. 
Every social system has a kind of feedback mechanism. Thus it is reasonable to expect 
that the quality of a product, initially released in 
the market, is also adjusted according to the ever upgraded opinions (preferences) of the
agents and the economic constraints. 

In this paper we examine a model which 
has been recently proposed by Solomon {\em et al.} \cite{solo} (also see ref.\cite{huang}) in a context of how 
an opinion is formed, when a commercial product is released in the market.
The product
is sometimes chosen by a spanning population of a region or sometimes 
fails to hit the target. 
Naturally, a concept of percolation \cite{stau} comes in and it has been termed 
as `{\it Social Percolation}' in ref.\cite{solo}).

The above `social percolation' model is based on a
social network which, for example,  may be a two dimensional regular network.
The agents are assumed to be situated on the sites of the 
network which provides the frame of fixed connectivity. They decide to buy 
some product (or watch a movie) if that qualifies according to their personal preferences. 
There is no other external influence in decision making.
Thus the `{\it quality}' and the personal `{\it preference}' are two variables in the 
model, the former is the global and the later being a local variable. 

The steps are the following: 

$\bullet$~To start with it is assumed that the agents along one 
boundary line are 
informed about the product. Each agent has a certain independent personal 
preference ($p_i$, for the $i$th agent) to begin with.
A uniform random distribution is chosen for this purpose so that $1 > p_i > 0$, for all $i$. The 
quality $q$ is assigned a certain value ($ q < 1$) at the beginning.\\

$\bullet$~Now if the quality is greater than the preference of an 
agent, $p_i < q$, the agent $i$ responds and the information is passed on 
to the nearest neighbour uninformed agents. Otherwise the agent remains inactive and the
information is not propagated to any of its nearest neighbours. 
After it responds, the agent $i$ enhances 
its personal preference from $p_i$ to $p_i + \delta p$. 
Else the inactive agent reduces its preference from 
$p_i$ to $p_i - \delta p$ at the next time step.
Those who remain uninformed in the process keep their preferences unchanged.\\

$\bullet$~At the end of one sweep (one time unit) of the whole system, 
it is checked if any of the agents on the other end has got the information 
and has responded. If so that means the information is propagated from 
one end to the other of the community.
That is, in the language of percolation, it is checked if the 
responded agents span or percolate the system from one end to the other.
This is a case of commercial success. So, in such a case, for
financial constraints, the quality is reduced from $q$ to $q - \delta q$ for
the next time step. 
Otherwise (in case of failure) the quality is increased 
from $q$ to $q + \delta q$.\\ 

The above three steps constitute the execution of the model of Solomon 
{\em et al.} which shows an interesting `self-organized' social 
dynamics \cite{soc}. A Leath-type algorithm is implemented to upgrade the
sites sequentially from a set of agents (sites) who are informed at the 
start.

In the above work \cite{solo} it is assumed that the population is 
distributed homogeneously on a social space and so the dynamics is 
tested on a regular square lattice.
Here in the present work we ask the question of how the social structure
plays a role in the social dynamics.
It may happen that the population is not homogeneously widespread
in a big locality. Because of geographical and other reasons, there may 
be certain clusterings of houses which may be connected by pathways. 
Thus there can be an entire inhomogeneous but correlated network of such 
population formed in a big region. 
Our interest is to examine how this correlated inhomogeneous 
structure plays role in the dynamics of the simple model system in contrast to 
the earlier studies \cite{solo} done on homogeneous random percolating network. 

For the purpose of this work, in order to generate correlated inhomogeneous 
network, we create spanning clusters of up spins at the 
critical temperature $J/K_B T_c= 0.44$ of a 2D Ising model at zero field
in the sense of Fortuin-Kasteleyn-Coniglio-Klein-Swendsen-Wang cluster 
definitions \cite{stau}. We removed bonds between two parallel (up) spins with
a probability $P =\exp(-2J/K_BT)$. For a typical $101\times 101$ and 
a $1001\times 1001$ systems we iterate up to $10^4$ Monte Carlo steps per spin.

Now, precisely, we check for a spanning cluster of neighbouring up spins 
connected by unremoved bonds. 
The agents are now situated on a tenuous and inhomogeneously connected 
network. 
Thus instead of random site percolation as in Solomon {\em et al.} we use 
Ising-correlated site-bond percolation to take into account the herding of
human beings. In the rest of the work we refer the homogeneous network
of ref.\cite{solo} as 'random case'.

Below we present our observations on various aspects of the dynamics and 
we discuss the related scenario.
For our entire work we chose $\delta p = 0.001$ and 
$\delta q= 0.001$ and all the quantities are calculated on typical networks of 
size $101\times 101$ and $1001\times 1001$. The initial value of $q$ was chosen 
to be 0.5. The late time dynamics does not depend on this choice.

We enumerate the time evolution of the quality $q$ and the average 
preference $<p> = (1/N)\sum\limits_{i=1}^{N} p_i$, where $N$ is number of 
agents on the spanning network. Both the quantities increase with time while $q$ 
shows a plateau type behaviour (not shown here) for an intermediate regime and $<p>$ increases initially 
in a power law. 
In Fig.1, we have plotted the difference of the above two, $q - <p>$, with time $t$.
To demonstrate a size effect we also plot the same for a network of size 
$500\times 500$ (indicated in the figure). 
For a comparison we plot the above quantity for the case of homogeneous
random case. The difference in the two cases is apparent: in the Ising correlated 
case the difference $(q - <p>)$ ever increases with time after a period of 
nonmonoticity, while in the random case the same goes down or remains constant in the
same time domain.
This reason may be attributed to the more frequent failures 
(no spanning) in this case, the quality 
of the product had to be increased while at the same time because of less 
effective communication through the tenuous network, many of the agents remain
uninformed and thus average preference, although it increases with time, 
remains low compared to $q$.

To check the time evolution of the distribution of the agents personal preferences, we plot the distribution of $p_i$'s for three 
different times, $t=200$ (early time), $t=1000$ (intermediate time) and at
$t=2500$ (a relatively late time) in Fig.2, Fig.3 and in Fig.4 respectively.
The value of $q$ is indicated in the figures in each case for a
guideline.
In all cases the average of $p_i$'s ($<p>$) is always less than $q$. 
In the early time the distribution is more or less uniform and soon the 
preferences start clustering around a value, a sharp peak appears and that 
shifts to the right. This reflects the fact that the agents are again and again
exposed to the information of the product so that they thrive for the 
high quality product and thus enhance their personal taste. 
However, apart from the sharp peak in the intermediate and late times, there 
appears to be a number of agents whose preferences remain far below. These are
counted in the average $<p>$ which brings down the value significantly 
from the most probable value and thus keeps it far below the quality ($q$) 
of the product provided at that time.
We have checked that this is even more prominent at very late times when
the distribution is clearly divided into two parts, one around the sharp 
peak on the right and a number of agents' preferences cluster around 0. 
Thus this further demonstrates the drifting of average preference away from the 
quality value, as time passes in the Ising correlated case. 
For comparison with the random homogeneous case Fig.5 is plotted corresponding 
to that in Fig.4. 
The left side of the peak in Fig.5 decays roughly exponentially and 
thus the contribution of the agents having such preference values to the 
average is much less in this case as compared to the case of Fig.4 for Ising 
correlated network.

We enumerate some additional quantities in order to 
understand the additional influences of the correlated 
inhomogeneous network on the dynamics.
We look at a magnetization-like quantity and compare 
our present case with that of the random percolation model. 
Let us define a quantity $d_i$, which we may call as decision of the agent $i$:
$d_i = +1$, if the agent $i$ chooses to buy the product or go for a movie 
($p_i < q$) and $d_i = -1$ otherwise ($p_i > q$).
The average decision, $<d> = (1/N)\sum\limits_{i=1}^N d_i$, in the model, interestingly,
oscillates in time between two well separated branches. Again for a comparison we plot the above for Ising correlated and for the random case, shown in Fig.6 and Fig.7 respectively. 
Although the qualitative behaviour is similar in two cases, there is a 
noticeable quantitative difference which may be important in determining 
some other quantities of interest. 
The average decision $<d>$ is always positive (high or low) as can be seen in Fig.6 whereas
the same attains negative values in the lower branch in the random 
case (Fig.7). This 
presumably tells us that a comparatively larger fraction of agents (due to the many smaller
non-spanning clusters) abstain from buying products at each time,
while in the Ising correlated case the fraction of agents who abstain from buying the product at each time step is comparatively less. However, as may be 
expected, there are a certain fraction 
of agents which are always difficult to reach due to ill connectivity (the structure of the
tenuous network) in Ising correlated case. Therefore, their preferences remain low 
compared to the most probable preferences in the social network. Now we see that this fact is
reflected further in other quantities we define below for brevity. 

Within the framework of the social percolation model one may define a concept of
`profit' (or `loss') and the accumulation of `wealth'.
Money spent by the producer/ manufacturer is proportional to the quality $q$ of
the product. Therefore, the price of the product could be proportional to 
$q/N$, where $N$ is the number of agents targeted. But the product is sold
among the agents whose preferences $p_i$ are less than the quality $q$ at that 
time. So we can define a quantity which we may call `{\it profit}' 
(the negative of this may be called '{\it loss}'): 
Profit = $(q/N)\sum\limits_{p_i < q} p_i - q$. 
The cumulative sum of the so-called profit (or loss) with time
may be thought of as `{\it wealth}'.

We numerically evaluate the above quantities in the random case (homogeneous structure) 
and later test those quantities for the inhomogeneous Ising correlated 
network.   
The distinct two branches shown in Fig.8 may be a company's profit (and loss) with time.
In Fig.9 we draw a distribution of these profit (or loss) values with time.
This shows a bimodal distribution. The 'wealth' is plotted in Fig.10 where we see that the
wealth is negative in the entire time domain shown but it tries to recover the total 
loss with time, the curves goes through a dip minimum and then a sharp rise. 

In Fig.11, Fig.12 and in Fig.13 we demonstrate the above
quantities (as indicated in the figure) for a Ising correlated network corresponding 
to Fig.8, Fig.9 and in Fig.10 respectively. 
Now the quantitative as well as a qualitative differences are apparent. In this type of
social network (under the scope of the present simple model) a company does not probably 
suffer much loss. Of course in this type of analysis the `price' is not fixed, neither the
quality or the preference values are bounded. This, however, indicates that the 
social structure (this means the connectivity among agents here) seems to play a 
definite role in the process of dynamics of any financial interest.

In conclusion, we may say that the indication of self-organized like behaviour
(as that found in the same model but for a random homogeneous lattice\cite{solo}) 
is probably absent here in the Ising correlated network. 
However, this correlated, inhomogeneous and tenuous network brings into 
the possibilities of many other interesting differences as compared to the random case.
The spread of information in all directions is what dictates the quantitative and in 
some cases qualitative change in some relevant quantities.

\vskip 0.5in
\noindent{\bf Acknowledgment:}
\vskip 0.2in

This work is supported by SFB 341 at the {\it Institute for Theoretical 
Physics, Cologne University}. Discussions with A. Novak and Z-F Huang 
are thankfully acknowledged.

\newpage

\begin{figure}[htb]
\centerline{\psfig{figure=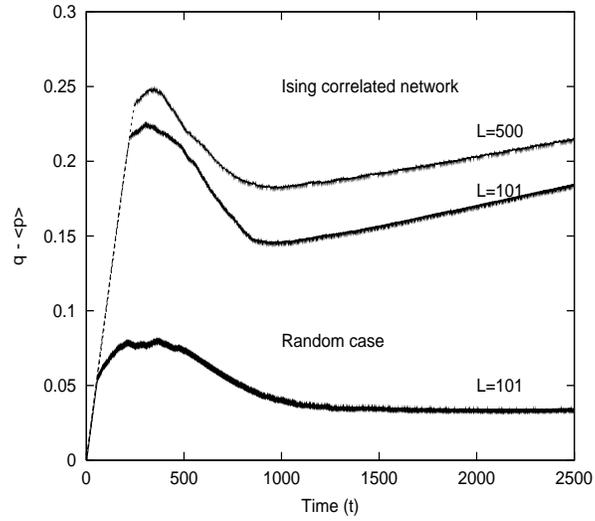, width=7cm, height=8 cm ,angle=-90}}
\caption{Time variations of the difference between the quality ($q$) of 
the product and the average preferences ($<p>$) of the agents are shown 
for Ising correlated network and that for the homogeneous network 
of random percolation case for comparison. To demonstrate the dependence on
system size one set of data is also plotted for a bigger system (indicated
in the figure) for the Ising correlated network.
}
\end{figure}

\begin{figure}[htb]
\centerline{\psfig{figure=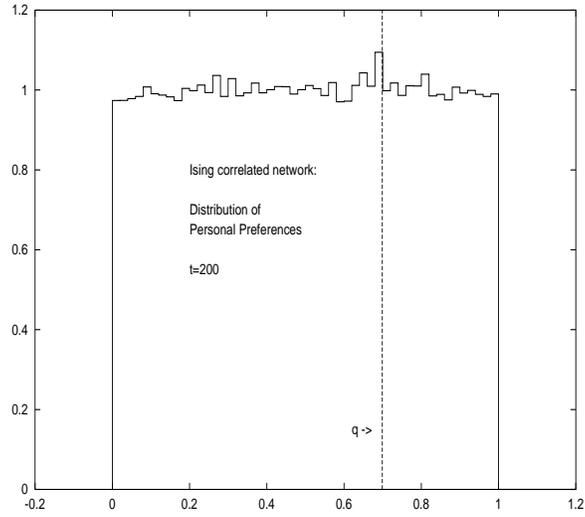, width=7cm, height=8 cm, angle=-90}}
\caption{Plotted here is the early time ($t=200$) distribution of personal 
preferences for a Ising correlated network on a $1001\times 1001$ lattice. The 
position of the value of the quality $q$ is indicated for comparison.}
\end{figure}

\begin{figure}[htb]
\centerline{\psfig{figure=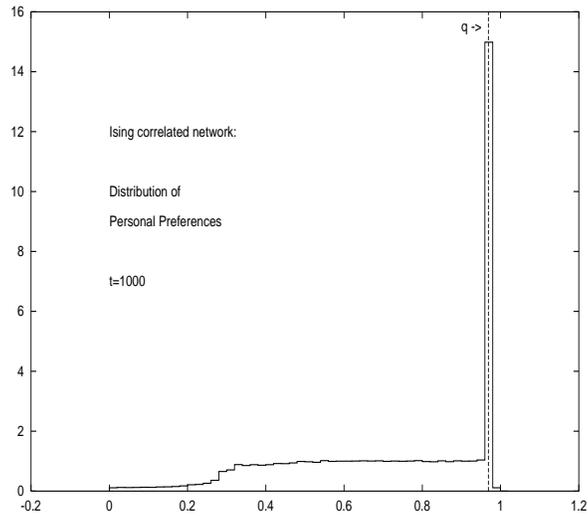, width=7cm, height=8 cm, angle=-90}}
\caption{Plotted here is the distribution of the same as that in Fig.2 but 
at a later time $t=1000$. }
\end{figure}

\begin{figure}[htb]
\centerline{\psfig{figure=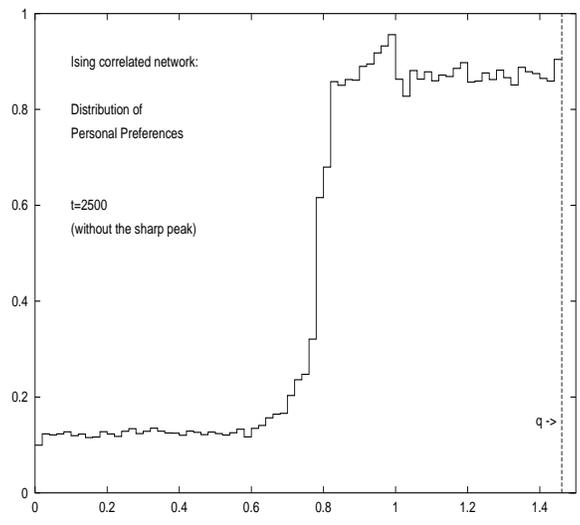, width=7cm, height=8 cm ,angle=-90}}
\caption{The distribution of personal preferences (on Ising correlated network)
at a late time $t=2500$. This is plotted here without the dominating 
sharp peak (right end). This shows the distribution profile for the 
preferences of numerous other agents than those clustered around the 
most probable value.}
\end{figure}

\begin{figure}[htb]
\centerline{\psfig{figure=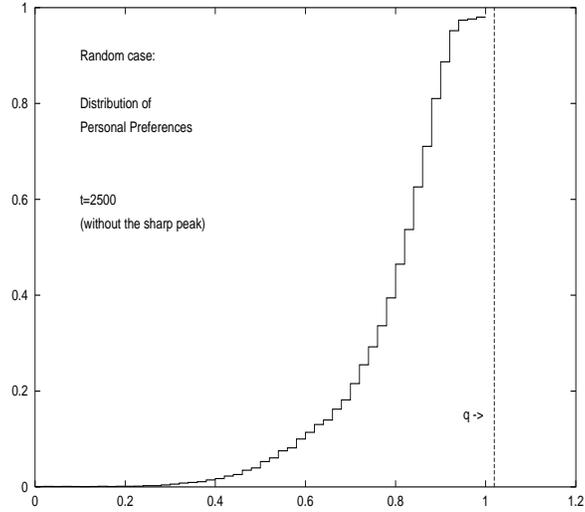, width=7cm, height=8 cm ,angle=-90}}
\caption{The distribution (without the sharp peak) of personal preferences
at $t=2500$ for a homogeneous network (random percolation case) for 
a $1001\times 1001$ lattice. This is plotted here to have a comparison 
with the case in Fig.4.}
\end{figure}

\begin{figure}[htb]
\centerline{\psfig{figure=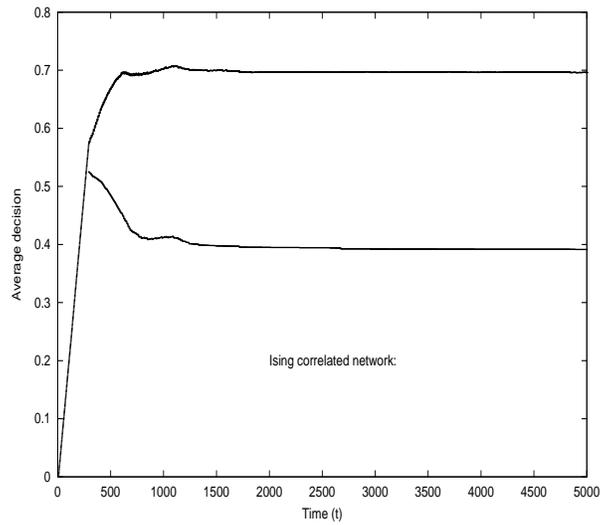, width=7cm, height=8 cm ,angle=-90}}
\caption{Average decision ($<d>$) with time ($t$) for the Ising correlated 
network of size $1001\times 1001$.}
\end{figure}

\begin{figure}[htb]
\centerline{\psfig{figure=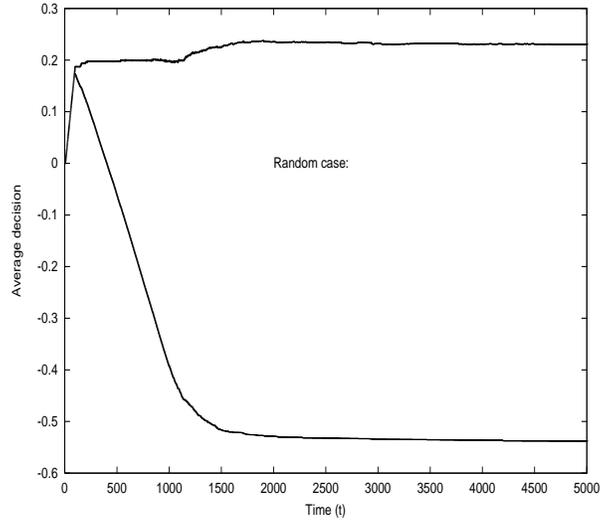, width=7cm, height=8 cm ,angle=-90}}
\caption{Average decision ($<d>$) with time ($t$) is plotted here for the
random homogeneous case for a $1001\times 1001$ lattice to be compared with 
Fig.6.}
\end{figure}

\begin{figure}[htb]
\centerline{\psfig{figure=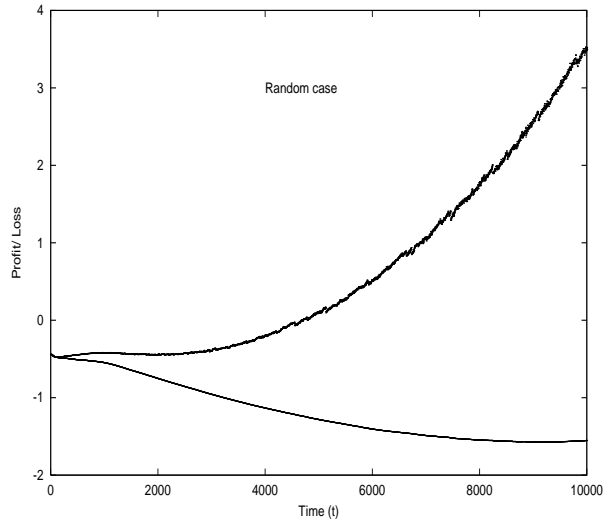, width=7cm, height=8 cm ,angle=-90}}
\caption{Time variation of `profit'/ `loss' in the random percolation case}
\end{figure}

\begin{figure}[htb]
\centerline{\psfig{figure=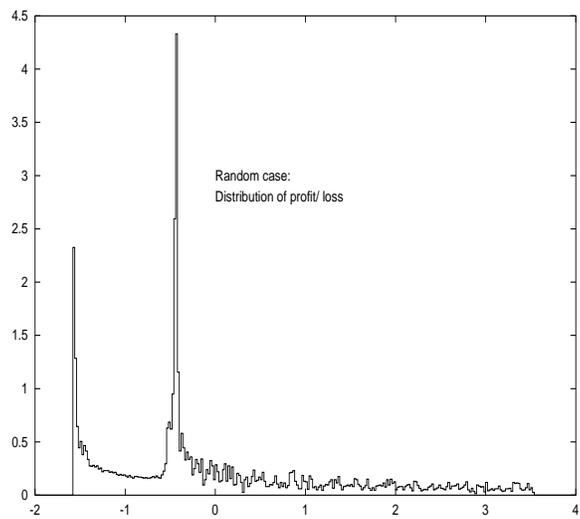, width=7cm, height=8 cm ,angle=-90}}
\caption{Distribution of `profit'/ `loss' in the random percolation case}
\end{figure}

\begin{figure}[htb]
\centerline{\psfig{figure=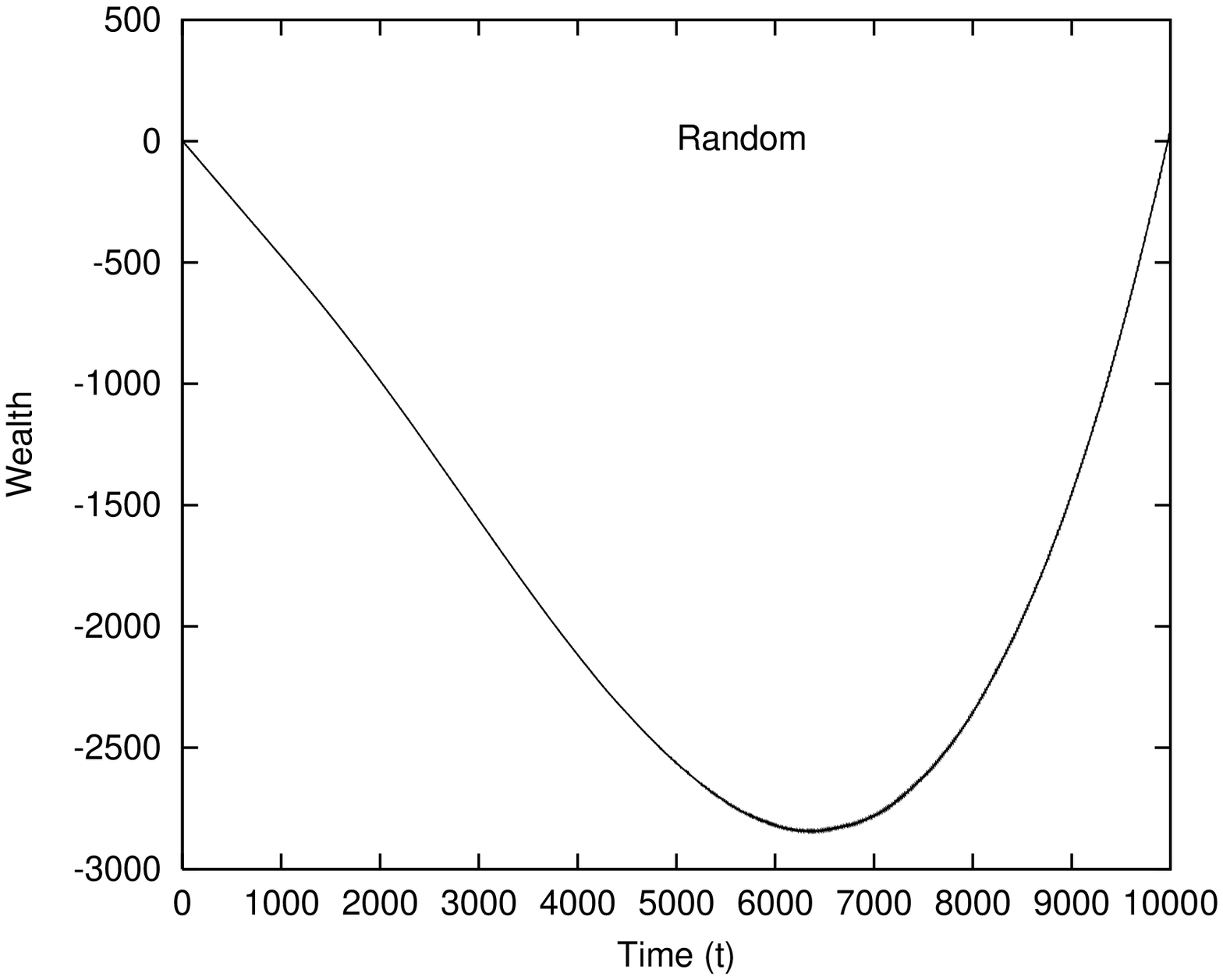, width=7cm, height=8 cm ,angle=-90}}
\caption{Accumulation of `wealth' in the random percolation case}
\end{figure}

\begin{figure}[htb]
\centerline{\psfig{figure=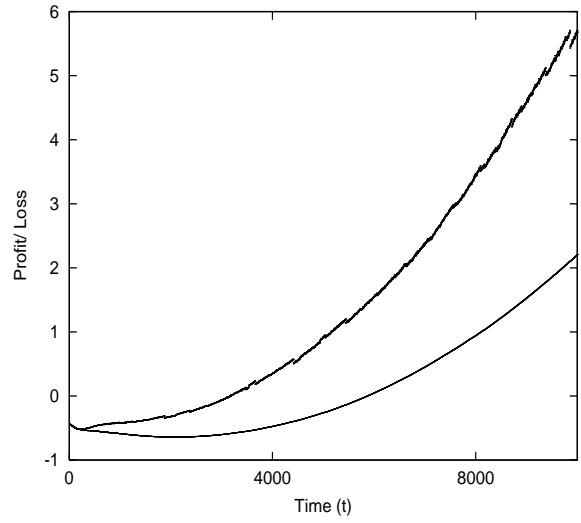, width=7cm, height=8 cm ,angle=-90}}
\caption{Time variation of profit/ loss in the Ising correlated case}
\end{figure}

\begin{figure}[htb]
\centerline{\psfig{figure=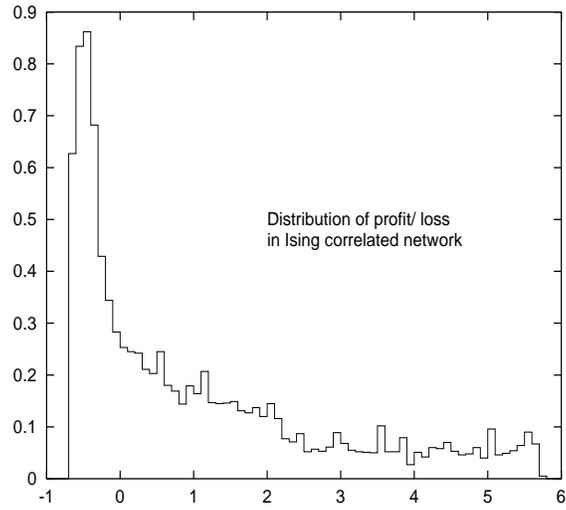, width=7cm, height=8 cm ,angle=-90}}
\caption{Distribution of profit/ Loss in the Ising correlated case}
\end{figure}

\begin{figure}[htb]
\centerline{\psfig{figure=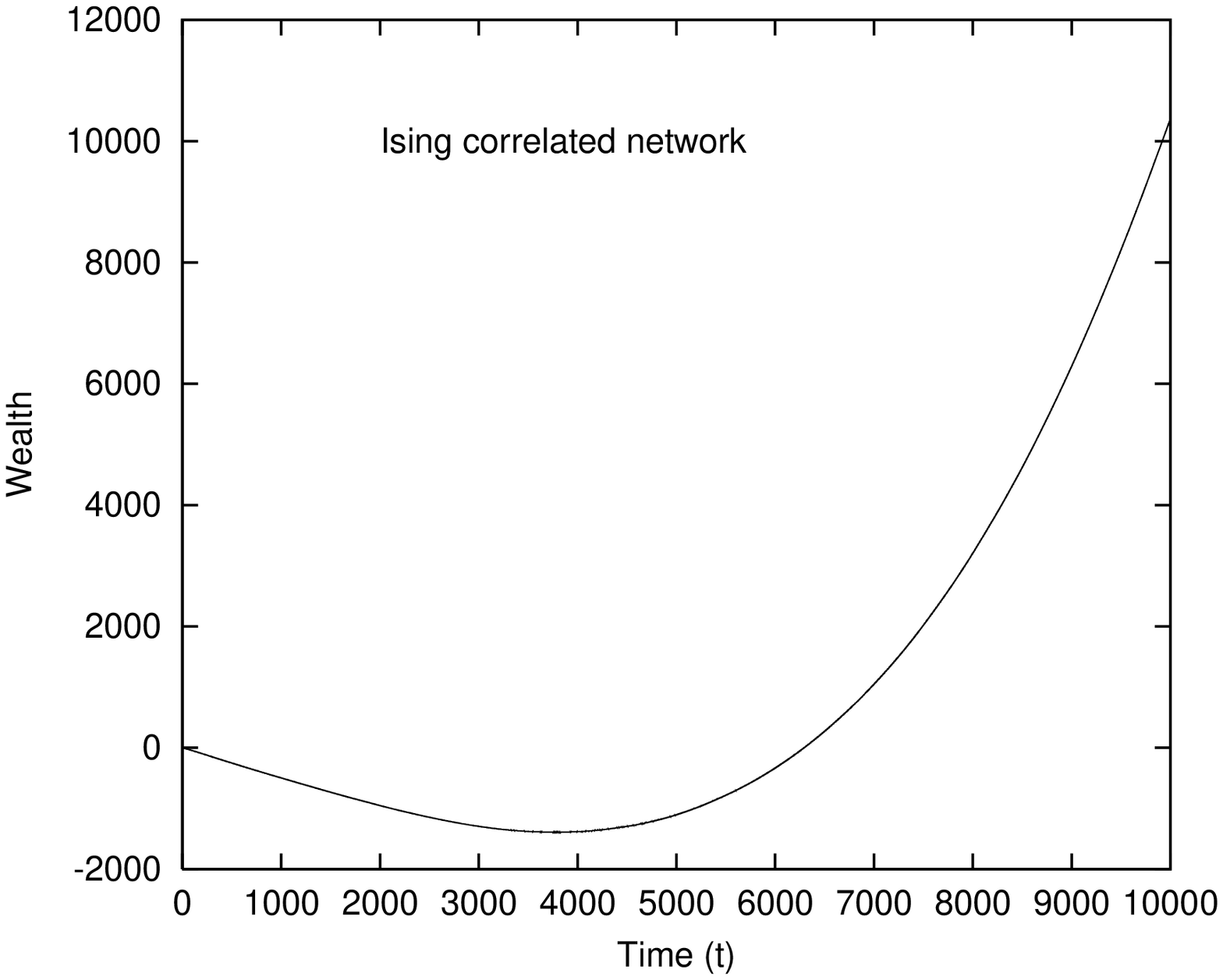, width=7cm, height=8 cm ,angle=-90}}
\caption{Accumulation of `wealth' in the random percolation case}
\end{figure}

\end{document}